\documentclass[%
 aip,
 amsmath,amssymb,
reprint,
floatfix]{revtex4-1}
\input epsf
\usepackage{lipsum}
\usepackage{latexsym}
\usepackage{color}
\usepackage{soul}
\usepackage{hhline}
\usepackage[normalem]{ulem}
\usepackage{graphicx}
\usepackage{psfrag}
\usepackage[nice]{nicefrac}

\DeclareMathOperator\erf{erf}
\DeclareMathOperator\erfc{erfc}
\usepackage{array}
\newcolumntype{M}[1]{>{\centering\arraybackslash}m{#1}}
\newcolumntype{L}[1]{>{\raggedright\arraybackslash}m{#1}}
\draft 

\begin{document}


\title{Voltage-Driven Translocation:~Defining a Capture Radius} 



\author{Le Qiao}
\author{Maxime Ignacio}
\author{Gary W. Slater}

\affiliation{University of Ottawa,Department of Physics, University of Ottawa, Ottawa, Ontario K1N 6N5, Canada}


\date{\today}

\begin{abstract}
Analyte translocation involves three phases: (i) diffusion in the loading solution; (ii) capture by the pore; (iii) threading. The capture process remains poorly characterized because it cannot easily be visualized or inferred from indirect measurements. The capture performance of a device is often described by a \textit{capture radius} generally defined as the radial distance $R^*$ at which diffusion-dominated dynamics cross over to field-induced drift. However, this definition is rather ambiguous and the related models are usually over-simplified and studied in the steady-state limit. We investigate different approaches to defining and estimating $R^*$ for a charged particle diffusing in a liquid and attracted to the nanopore by the electric field. We present a theoretical analysis of the P\'{e}clet number as well as Monte Carlo simulations with different simulation protocols. Our analysis shows that the boundary conditions, pore size and finite experimental times all matter in the interpretation and calculation of $R^*$. 
\end{abstract}


\maketitle 

\section{Introduction}
\label{section:intro}
The field-driven translocation of a charged analyte across a nanopore~\cite{kasianowiczCharacterizationIndividualPolynucleotide1996,mellerVoltageDrivenDNATranslocations2001,leeElectrophoreticCaptureDetection2004,fologeaSlowingDNATranslocation2005,heControllingDNATranslocation2011,pastoriza-gallegoDynamicsUnfoldedProtein2011,mihovilovicStatisticsDNACapture2013a,mirigianTranslocationHeterogeneousPolymer2012,palyulinPolymerTranslocationFirst2014a,briggsDNATranslocationsNanopores2018,beamishIdentifyingStructureShort2017,seanLangevinDynamcisSimulations2017,charronPreciseDNAConcentration2019} has been studied extensively over the past two decades due to its potential applications, especially for DNA detection and sequencing.  Translocation involves three major steps for the analyte: (i) diffusion towards the nanopore; (ii) capture by the electric forces near the nanopore; (iii) passage (or threading for polymers) through the nanopore. The last step, which has been the subject of numerous \textit{in vivo} and \textit{in silico} studies, is controlled by factors such as the electro-osmotic flow\cite{wongPolymerCaptureElectroosmotic2007}, pressure and concentration gradients\cite{wanunuElectrostaticFocusingUnlabelled2010,jeonPolymerCaptureAhemolysin2014,hatloTranslocationDNAMolecules2011,jeonElectrostaticControlPolymer2016}, entropic effects\cite{muthukumarTheoryCaptureRate2010b,katkarRoleNonequilibriumConformations2018}, and electric forces\cite{mellerVoltageDrivenDNATranslocations2001,wanunuElectrostaticFocusingUnlabelled2010}. However, the diffusion-and-capture process remains rather ill-understood because it cannot easily be observed; moreover, in the case of polyelectrolytes, the interplay between the diffusion/drift and the polymer conformational deformations complicates modelling\cite{vollmerTranslocationNonequilibriumProcess2016,farahpourChainDeformationTranslocation2013a}.

Ideally, a translocation device should have a large capture zone and a high capture rate\cite{seanLangevinDynamcisSimulations2017,waughInterfacingSolidstateNanopores2015,hatloTranslocationDNAMolecules2011}. The former is generally described by a quantity called the capture radius, $R^*$, which is typically reported~\cite{chenProbingSingleDNA2004} to be $\sim\mu m$. Theoretically, $R^*$ is loosely defined as the radial distance at which diffusion-dominated dynamics (at large distances) crosses over to drift-dominated dynamics (close to the pore) in a steady-state regime. 

One way to estimate $R^*$ for a particle is thus to compare its electrostatic and thermal energies~\cite{chenProbingSingleDNA2004,nakaneEvaluationNanoporesCandidates2002a,gershowRecapturingTrappingSingle2007a,grosbergDNACaptureNanopore2010a,wanunuElectrostaticFocusingUnlabelled2010,nomidisDNACaptureClyA2018a}. The minimal work needed to bring it from $r$ to infinity is\cite{grosbergDNACaptureNanopore2010a}
\begin{equation}
\label{Work}
\frac{W(r)}{k_BT}= \frac{\mu }{D}~V(r) ,
\end{equation}
where $D$ and $\mu$ are the particle's diffusion coefficient and electrophoretic mobility, respectively, $k_BT$ is the thermal energy, and
\begin{equation}
\label{eq:pointV}
V(r) = \Delta V ~ \frac{r_e}{r}
\end{equation}
is the point-charge approximation\cite{wanunuElectrostaticFocusingUnlabelled2010,grosbergDNACaptureNanopore2010a,nomidisDNACaptureClyA2018a,rowghanianElectrophoreticCaptureDNA2013a} for the electric potential at a distance $r$ when a voltage difference $\Delta V$ is applied across the system. Here $r_e=r_p/(\frac{2l}{r_p}+{\pi})$ is the characteristic length of the electrostatic potential outside a pore of radius $r_p$ and length $l$ \cite{nomidisDNACaptureClyA2018a}. The capture radius is then defined as the distance $R^*$ at which $W(R^*) \! = \! k_BT$ \cite{wanunuElectrostaticFocusingUnlabelled2010,grosbergDNACaptureNanopore2010a,rowghanianElectrophoreticCaptureDNA2013a}. Using eq.~\ref{Work}, this condition reads $V(R^*)= D/\mu$. Given eq.~\ref{eq:pointV}, the resulting capture radius is
\begin{equation}
\label{R*1}
R^*= \frac{\mu \Delta V }{D}~ r_e.
\end{equation}

Alternatively, we can define the capture radius as the point of no return\cite{nomidisDNACaptureClyA2018a,wanunuElectrostaticFocusingUnlabelled2010,nakaneEvaluationNanoporesCandidates2002a} (or \textit{event horizon} to borrow an expression from General Relativity), \textit{i.e.} the distance below which the particle cannot escape the attraction of the pore. The field-driven flux at $r$ is simply
 \begin{equation}
\label{EqEflux1}
J_E(r)= \mu(r)E(r) C(r),
\end{equation}
where $E(r)\!=\!\mathrm{d}V(r)/\mathrm{d}r$ is the electric field and $C(r)$ is the particle density. In this approach, the hemisphere of radius $R^*$ is treated as a perfect absorber and the electric field is neglected beyond. The diffusive flux of particles reaching this hemisphere from the bulk is\cite{grosbergDNACaptureNanopore2010a,nakaneEvaluationNanoporesCandidates2002a}~
\begin{equation}
\label{flux}
J_D=-D C_o/R^*, 
\end{equation}
where $C_o \! \equiv \! C(\infty)$ is the density in the undisturbed far bulk. The capture radius $R^*$ is then defined as the location where the two fluxes are equal: $J_D(R^*) \! = \! J_E(R^*)$. To complete the calculation we assume that $C(R^*)=C_o$ in eq.~\ref{EqEflux1}; we then recover the result in eq.~\ref{R*1}.  

In a third approach, $R^*$ is defined as the distance at which the times taken by a particle to find the nanopore solely by diffusion or solely by electrophoresis are equal (this is basically a P\'eclet number method)~\cite{grosbergDNACaptureNanopore2010a}. A simple scaling argument suggests that the diffusion time is 
\begin{equation}
\label{tauDscaling}
\tau_D^o \approx {R^*}^2/D.
\end{equation}
The drift time over distance $R^*$ is approximated as
\begin{equation}
\label{tauE}
\tau_E^o \approx {R^*}/{\mu E(R^*)}.
\end{equation}
If we use the expression for $E(r)$ given before, we again obtain eq.~\ref{R*1}. However, the scaling argument for $\tau_D^o$ does not include the size $r_p$ of the target that has to be found by diffusion: $\tau_D^o$ is thus underestimated. Moreover, since the electric field decays with $r$, the drift time should be integrated over the field lines instead of using the field at $R^*$ in eq.~\ref{tauE} -- in other words, $\tau_E^o$ is overestimated. In conclusion, eq.~\ref{R*1} provides a lower bound for $R^*$.

In this paper, we revisit some theoretical approaches and present 2D Lattice Monte Carlo (LMC) methodologies and results with different simulation protocols and boundary conditions in order to propose and test various definitions of $R^*$ for point-like analytes. We also examine the impact of the fields lines near the pore, especially for short times.

\section{Basic elements and estimates}

After looking at typical experimental conditions and the field lines around a nanopore, this section presents an analysis of the key variables, time and length scales.

\subsection{Experimental and simulation parameters}
\label{section:parameter}
Our theoretical and computational studies focus on the capture process in the translocation system shown in Fig.~\ref{Fig_system}. Although the analyte is point-like, its properties are chosen to match that of DNA (see below) in order to compare to available experimental data (while this is not our main goal, such comparisons can be useful). We neglect DNA conformational entropic effects, analyte-analyte interactions, hydrodynamic interactions with the walls and electroosmotic flow. Finally, the pore is treated as a perfect absorber. We use the pore radius $r_p$ as the unit of length and $\tau_o \! = \! r_p^2/D$ as the unit of time.

\begin{figure}
\centering
\includegraphics{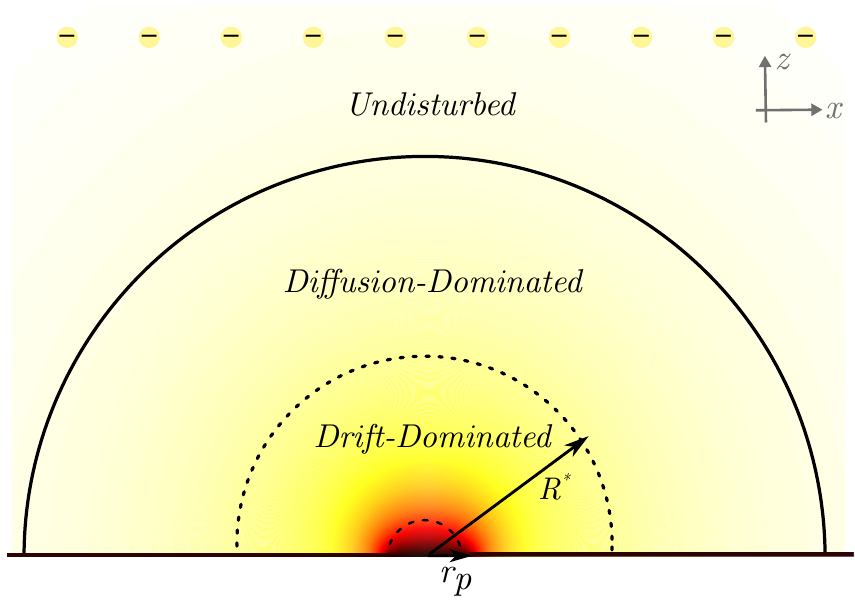}
\caption{A schematic view of a nanopore capture system: $r_p$ is the pore radius and the background colour codes for the electric field strength (higher fields in red). The dashed lines depict the capture radius  $R^*$ and a hemisphere of radius $r_p$. The solid line depicts the boundary of the region that remains undisturbed during an experiment.}
\label{Fig_system}
\end{figure}

The maximum electrostatic energy of a particle of charge Q is $\psi_o= Q \times \Delta V$. For a polyelectrolyte like DNA, the relevant electrophoretic charge is $Q=k_BT \mu /D$ because the counterions, which move in the direction opposite to the DNA in an electric field, affect the local friction drag on the DNA backbone (this leads to a well-known failure of the fluctuation-dissipation theorem -- see, for example, ref~\cite{grosbergDNACaptureNanopore2010a,longSimultaneousActionElectric1996}). Equation~\ref{R*1} can thus be rewritten as  
\begin{equation}
\label{eq:R*2}
R^*= \frac{\psi_o}{k_BT}~ r_e \equiv \lambda_e.
\end{equation}
This basic capture radius is called the electrostatic length $\lambda_e$ in this paper as it is the radial distance at which $W(r \! = \! \lambda_e) \! = \! k_BT$, and it will be used as a measure of the field intensity.

We will mostly use values for a 250 base ssDNA molecule\cite{nkodoDiffusionCoefficientDNA2001a}: $D \! \approx \! 17~\mu m^2/s$ and $\mu \approx 4.1 \times 10^{4}~\mu m^2/Vs$, giving $Q \approx 60\ e$ ($\approx \nicefrac{1}{4}$ of the nominal charge). With a voltage $\Delta V \! = \! 800~mV$, the scaled potential energy is $\psi_o/k_BT \approx 1800$.  Using a pore radius $r_p \! = \! 5~nm $ and a pore length $l\!=\!2r_p$,  we have $r_e=r_p/(4+\pi)$ and $\lambda_e \approx 250 r_p \approx 1~\mu m$. The time unit is thus $\tau_o =r_p^2/D \approx 1.5 ~\mu sec$ here; a typical experimental duration of $t_{exp}=0.5 ~h$ is then $ \approx 10^9 \tau_o$.

\subsection{The electric field}
In most theoretical studies~\cite{grosbergDNACaptureNanopore2010a,chenProbingSingleDNA2004,nakaneEvaluationNanoporesCandidates2002a,wanunuElectrostaticFocusingUnlabelled2010,gershowRecapturingTrappingSingle2007a,rowghanianElectrophoreticCaptureDNA2013a}, the electric field is modelled by treating the nanopore as a point charge, as described by eq. \ref{eq:pointV}, a good approximation if $r \! \gg \! r_p$. For distances comparable to $r_p$, though, the spherical symmetry is broken and this approximation is unreliable. Instead, one can solve the Laplace's equation in oblate coordinates $(\xi,\eta,\phi)$ for the electric field outside the pore\cite{kowalczykModelingConductanceDNA2011,farahpourChainDeformationTranslocation2013a}: 
\begin{equation}
 \label{eq:sol}
  V(\xi) = \tfrac{2 }{\pi}~ \delta V ~\arctan \left[\sinh(\xi)\right],
\end{equation}
where $\xi \in (- \infty, + \infty)$ is the oblate distance to the pore and $\delta V$ is the potential drop from infinity to pore. The prefactor can be written as
\begin{equation}
\label{eq: pdeop}
\frac{2}{\pi}~ \delta V=  \frac{r_e}{r_p}~\Delta V,
\end{equation} 
thus establishing a connection between $\delta V$ and $\Delta V$.
Note that close to the pore ($\xi < 1$ or, equivalently, $r<r_p$), the field is actually rather flat -- see Appendix~\ref{app:Electric_field}. As $R^* \! \gg \! r_p$ for realistic cases, we will use the point-charge approximation for analytical calculations (indeed, the prefactor $\Delta V r_e/r_p^2$ in eq.~\ref{eq:Electric_zero_ap} is the value predicted by eq.~\ref{eq:pointV} at $r=r_p$). However, the flatness of the field near the pore will play a role in some cases.

\subsection{Preliminary considerations}
\label{sec:preliminary}

\noindent \textbf{Velocity}: The definition of $\lambda_e$ allows us to rewrite the radial drift velocity $v(r)=\mu E(r)$ as
\begin{equation}
 \label{Eqv_p2}
 v(r)=-{\lambda_e D}/{r^2}.
\end{equation}
Note that eq.~\ref{flux} is equivalent to a diffusive velocity 
\begin{equation}
 \label{Eqv_D2}
 v_D(r)=-{ D}/{r}.
\end{equation}
As expected, these velocities are equal at $r=\lambda_e$.\\

\noindent \textbf{Deterministic drift times}: A useful parameter is the field-driven deterministic time to drift from position $r_o$ to $r<r_o$ under the action of the electric field:
\begin{equation}
\label{tauEro2r}
    \tau_E(r_o,r)=\int_{r_o}^{r} \frac{\mathrm{d}r^{\prime}}{v_d(r^{\prime})}= \frac{ r_o^3-r^3}{ 3\lambda_e D}.
\end{equation}
For instance, the \textit{cleanup time} $\tau_\lambda$ needed to empty the capture radius zone $\lambda_e$ is then
\begin{equation}
\label{tauEL20}
    \tau_\lambda = \tau_E(\lambda_e,0)=\lambda_e^2/3 D,
\end{equation}
which is \nicefrac{1}{3} of $\tau_E^o$ in eq.~\ref{tauE}. However, eq.~\ref{tauEro2r} is using the point-like charge approximation and thus fails to take into account the fact that the field is fairly flat near the pore. The time to reach the pore from $r=r_p$ is
\begin{equation}
\label{tauErp2r}
    {\tau} _p \approx \frac{r_p}{|v(r_p)|} = \frac{r_p^3}{ \lambda_e D}=3 \tau_E(r_p,0).
\end{equation}
Equation \ref{tauEro2r} thus underestimates (by a factor of $\approx 3$) the time taken for the last part of the capture process. Since
\begin{equation}
    \label{eq:taurpovertaulambda}
    {{\tau}} _p/\tau_\lambda =3~(r_p/\lambda_e)^3 \ll 1,
\end{equation}
this has no impact in realistic cases.\\

\noindent \textbf{The capture rate in 3D}: The mean capture rate during the cleanup time $\tau_\lambda$ is 
\begin{equation}
    \label{MeanRate}
    \overline{\rho}_\lambda \approx \tfrac{2 \pi}{3} \lambda_e^3 C_o/\tau_\lambda=2 \pi D \lambda_e C_o~,
\end{equation}
while the initial rate (during the period required to "empty" the region of size $r_p$) is
\begin{equation}
    \label{InitialRate3D}
    \overline{\rho}_p \approx \tfrac{2 \pi}{3} r_p^3 C_o/{\tau}_p=\tfrac{1}{3}\overline{\rho}_\lambda.
\end{equation}
We thus expect the mean capture rate to initially (and rapidly) increase (by a factor $\approx 3$) because of the special field lines near the pore (eq.~\ref{eq:Electric_zero_ap}). More generally, the total number of particles captured during a period of duration $t$ is given by
\begin{equation}
    \label{Eq:sumcapture}
    N(t) \approx \tfrac{2}{3}\pi r^3(t)C_o,
\end{equation}
where $r(t)$ is the size of the region emptied at time $t$. The capture rate at time $t$ is thus
\begin{equation}
    \label{Eq:instRate}
    \rho(t) ={\partial N(t)}/{\partial t}=2\pi C_o r^2(t)\frac{\mathrm{d}r(t)}{\mathrm{d}t}.
\end{equation}
If we use eq.~\ref{tauEro2r} to estimate the size of this region, we obtain $r(t) \approx (3Dt\lambda_e)^{1/3}$ from which eq.~\ref{Eq:instRate} predicts that $\rho(t) \approx \overline{\rho}_\lambda$ would be constant at long times. \\

\noindent \textbf{The capture rate in 2D}: Although our work is about the physics of 3D systems, our simulations will be done in 2D, for efficiency reasons (see Section \ref{section:LMCm} for details). In this context, 2D means a slice of 3D system (as opposed to a system between two walls); therefore, a 3D electric field is used in both our analytical calculations and LMC simulations. In two dimensions, eq.~\ref{MeanRate} must be replaced by
\begin{equation}
    \label{MeanRate2D}
    \overline{\rho}_\lambda \approx \tfrac{\pi}{2} \lambda_e^2 C_o/\tau_{\lambda}=\tfrac{3\pi}{2} D C_o,
\end{equation}
where we used eq.~\ref{tauEL20} . Interestingly, the mean cleanup rate is not predicted to be a function of the field intensity in 2D. Similarly, the initial rate is now predicted to be
\begin{equation}
    \label{InitialRate2D}
    \overline{\rho}_p \approx \tfrac{\pi}{2}r_p^2 C_o /\tau_p = \tfrac{\pi \lambda_e}{2r_p} D C_o =  \tfrac{\lambda_e}{3 r_p}~ \overline{\rho}_\lambda.
\end{equation}
A large drop of $\overline{\rho}$ is thus predicted in 2D (instead of the increase predicted for 3D in the previous section). Unlike the 3D case, the instantaneous capture rate does not become time-independent in 2D: for instance, using 
\begin{equation}
    \label{N2D}
N(t) =\tfrac{1}{2}\pi r^2(t)C_o,
\end{equation}
we now find that 
\begin{equation}
    \label{InstantRate2D}
    \rho(t)=\tfrac{2}{3} \overline{\rho}_\lambda \times \left[ \tau_\lambda/t \right]^{1/3},
\end{equation}
from which we recover $\overline{\rho}_\lambda=\int_{0}^{\tau_\lambda} \rho(t) dt/\tau_\lambda = \rho_\lambda$, as it should. From eqs.~\ref{MeanRate2D} and \ref{InstantRate2D}, we also predict that $\rho(\tau_\lambda)=\pi D C_o$ should be field-independent.
\\

\noindent \textbf{Experiments:} In practice, the experimental duration $t_{exp} \gg \tau_\lambda$: the $\lambda_e$ zone is thus emptied $N\!=\!t_{exp}/\tau_\lambda \! \gg \! 1$ times during an experiment. If we assume that the concentration remains $C_o$ in this zone during the experiment, the upper bound for the total number of molecules captured is $M=N \times \frac{2 \pi}{3} \lambda_e^3 C_o=t_{exp} \times {2 \pi} D \lambda_e C_o=\overline{\rho}_\lambda t_{exp}$. Similarly, using the diffusion flux eq.~\ref{flux}, the number of molecules that have diffused into the nominal capture zone during the experiment is also  $M= 2 \pi  \lambda_e^2 |J_D|\times t_{exp}=\overline{\rho}_\lambda t_{exp}$. Therefore, $\overline{\rho}_\lambda$ is indeed the mean capture rate predicted by the basic theory described in the Introduction; moreover, the calculation also predicts a constant capture rate passed the initial jump mentioned above.
\\

\noindent \textbf{Numerical values}: We now look at numerical estimates for the case described previously. With $\lambda_e \! \approx \! 1 ~\mu m$, eqs~\ref{Eqv_p2} \! -- \! \ref{MeanRate} give $v(r_p)\! \approx \! 1~m/s$, $v(\lambda_e) \approx 17~\mu m/s$, $\tau_\lambda \approx 20~ms$ and $\overline{\rho}_\lambda \approx (4.7~\mu m)^3 C_o/s = (C_o/16~pM)/s$, which provides guidance for the concentration $C_o$ to be used in practice.

\section{Theoretical Approaches Revisited} 
\label{section:Theory}

\subsection{The P\'eclet argument with a small target}
\label{section:Peclet}

The dimensionless P\'{e}clet number ($Pe$) is used in Separation Science to characterize the competition between advection and diffusion.\cite{kirbyMicroNanoscaleFluid2010} Here we define $Pe$ as 
\begin{equation}
 \label{Pe1}
  Pe(r) = {\tau_D\left(r,0\right)}/{\tau_E\left(r,0\right)},
\end{equation}
where $\tau_D(r,0)$ is the mean time needed by a particle to find the pore (at $r=0$) by diffusion alone when starting from a radial distance $r$, and $\tau_E(r,0)=r^3/3D \lambda_e$ is the time for the same particle to travel to the pore when driven solely by the applied field -- see eq.~\ref{tauEro2r}.

The mean first passage time (MFPT) for a particle in Fig.~\ref{Fig_system} is analogous to a classical diffusion problem where a particle is initially located between two concentric spheres of sizes $R_b$ and $r_D \ll R_b$, with the internal one being absorbing and the largest one being reflecting\cite{rednerGuideFirstpassageProcesses2001,bergRandomWalksBiology1993,szaboFirstPassageTime1980a}. If the particle starts at radial position $R_b \ge r \ge r_D$, the MFPT is:
\begin{equation}
 \label{taud3D}
  \tau_D(r,r_D) = \frac{R_b^2}{6D}\left[\frac{2R_b}{r_D}\left(1 \! - \! \frac{r_D}{r}\right) + {\left(\frac{r_D^2-r^2}{R_b^2} \right)} \right].
\end{equation}
The position $r_D$ must be very close to the pore; we will use $r_D \!=\! r_p$ for simplicity. Using eqs. \ref{tauEro2r}, \ref{Pe1} and \ref{taud3D}, we can calculate $Pe(r)$; since $r_p \ll R_b$ and $r_p \ll R^*$ in practice, we obtain
\begin{equation}
 \label{Pe3d}
  Pe(r) \approx {R_b^3\lambda_e}/{r^3 r_p}~.
\end{equation}
The P\'{e}clet capture radius is the solution of $Pe(R^*) \! = \! 1$:
\begin{equation}
 \label{R*vsRb}
 R^* \approx R_b \! \times \! (\lambda_e/r_p)^{1/3}.
\end{equation}
Since $\lambda_e \! \gg \! r_p$, $R^*$ exceeds the box size $R_b$. However, both $\tau_E(R_b,0)$ and $\tau_D(R_b,r_p)$ also exceed any realistic experimental time. Therefore, our P\'{e}clet result actually implies that $R^*$ steadily increases during an experiment (\textit{i.e.}, the process is field-driven throughout). The time-dependent effective capture radius $R^*(t)$ is then obtained by inverting $\tau_E(r,0)=r^3/3D \lambda_e$:
\begin{equation}
 \label{Driftdistance}
{R^*(t)} \approx \left(3\lambda_e D t \right)^{1/3}=\lambda_e \times \left[t/\tau_\lambda\right]^{1/3}. 
\end{equation}
It increases as $\sim t^{\nicefrac{1}{3}}$ and exceeds $\lambda_e$ for times $t > \tau_\lambda$. 
For the previous example, with an experimental time $t_{exp}\!=\!0.5~h$ and a cleanup time $\tau_\lambda=20~ms$, the final capture radius is $R^*(t_{exp}) \approx 45 \lambda_e=45 ~\mu m$, almost two orders of magnitude larger than what eq.~\ref{eq:R*2} predicts. 

Interestingly, the situation is different in two dimensions since the MFPT is then
\begin{equation}
 2D:~ \tau_{D}(r,r_D) = \frac{R_b^2}{4D}\left[2 \ln\left({\frac{r}{r_D}}\right) + \left(\frac{r_D^2-r^2}{R_b^2} \right) \right].
\end{equation}
Combining this with eq.~\ref{Pe1} we obtain
\begin{equation}
 \label{eq:Pe2d}
2D:~Pe(r) \approx \frac{3R_b^2 \lambda_e}{2r^3} \left[ \ln\left({\frac{r}{r_D}}\right) \! - \! \frac{1}{2} {\left(\frac{r}{R_b} \right)}^2 \right] \!.
\end{equation}
To leading order, the solution of the P\'eclet condition $Pe(R^*)=1$ thus gives a length scale
\begin{equation}
    \label{eq:2DR*}
   2D:~ R^*\!\approx
\! (R_b^2\lambda_e)^{1/3},
\end{equation}
which means that we now have $R^* < R_b$. The time to reach the pore from this distance is $\tau_E(R^*,0) \approx R_b^2/3D = (R_b/\lambda_e)^2 \times \tau_\lambda$. If $t_{exp}$ is shorter than this time, eq.~\ref{Driftdistance} remains valid. Otherwise, we expect a transition to a diffusion-limited regime at long times.

\subsection{A P\'eclet-inspired flux method}

\label{section:PecletFlux}

In the flux method used in the Introduction, the hemisphere of radius $R^*$ is treated as an absorbing boundary (\textit{i.e.}, $C(R^*) \! = \! 0$) to obtain the diffusion flux, eq.~\ref{flux}. Yet, $C(R^*) \! = \! C_o$ is used in the calculation of the field driven flux, eq. \ref{EqEflux1}. This inconsistency cannot easily be fixed. Moreover, since both the flux and energy methods give $R^* \! = \! \lambda_e$, the fact that $W(\lambda_e) \! = \! k_BT$ implies that the $R^*=\lambda_e$ hemisphere is \textit{not} perfectly absorbing. One could try to get around this issue by using an hemisphere of radius $r_o$, with $r_p\leq r_o <\lambda_e$, as the location of the no-return point instead. As the field is extremely strong at $r_o=r_p$ (we have $W(r_p)\approx 250 ~k_BT$ for our example), this would truly be an absorbing boundary. However, any position $r_o$ for which $W(r_o) \gg k_BT$ can also be chosen. Unfortunately, using this approach to recalculate the two sides of the equality $J_D(R^*)=J_E(R^*)$ (the basis of the original flux method) fails to provide physically meaningful results (not shown). 

An alternative approach is to use "pure adsorption fluxes" instead of the "pure" adsorption times $\tau_D$ and $\tau_E$. We first compute the flux $J_E(r) \! = \! C(r,t)v(r)$ at a radial distance $r$ with the initial condition $C(r,0) \! = \! C_o$ and no diffusion. Since the molecules at position $r$ at time $t$ must all come from $r_o(r,t)$ at time $t \! = \! 0$, we have
\begin{equation}\label{Eq_driftC}
    C_E(r,t)=C(r_o,0)\times \left(\frac{r_o}{r}\right)^{2}=C_o \left(\frac{r_o}{r}\right)^{2}.
\end{equation}
We can obtain the initial position directly from eq.~\ref{tauEro2r}:
\begin{equation}\label{Eq_drift_r0}
    r_o(r,t)=r \times \left(1+ 3\lambda_e D  t/r^3\right)^{1/3}.
\end{equation}
The time-dependent concentration is thus given by
\begin{equation}
    C_E(r,t)=C_o\left(1+{3 \lambda_e Dt }/{r^3}\right)^{2/3},
\end{equation}
and the zero-diffusion flux $J_E(r,t)=v(r) C_E(r,t)$ is  
\begin{equation}
\label{eqJEo}
    J_E(r,t)=-  \frac{C_o D \lambda_e}{r^2}  \left[1+\frac{t/\tau_\lambda }{(r/\lambda_e)^3}\right]^{2/3}.
\end{equation}
Without diffusion, there is no steady-state and $J_E(r,t)$ increases with time since the electric forces concentrate the molecules in a smaller volume closer to the pore. 

To complete the calculation we need the time-dependent solution\cite{bressloffStochasticProcessesCell2014a} of the diffusion equation in absence of an external field but with an absorber at $r_D=r_p$: 
\begin{equation}
 \label{Eqdensityt1}
C(r,t) =C_o\left(1-\frac{r_p}{r}\right)+\frac{r_p C_o}{r}\erf{\left(\frac{r-r_p}{\sqrt{4Dt}}\right)}.
\end{equation}
Since $J_D =-D\frac{\partial C(r,t)}{\partial r}$, the diffusive flux is
\begin{equation}
J_D(r,t) = -\frac{ C_o D r_p }{{r}^2}   \left[ \erfc\left[{\delta}\right]+ \frac{r}{\sqrt{\pi D t}} ~ e^{ -\delta^2}\right]\!,
\label{JDflux3d}
\end{equation}
where $\delta=\delta(r,t)\equiv(r-r_p)/\sqrt{4Dt}$. 

We can now determine at which location $R^*(t)$ these two fluxes are equal by solving the relation $J_E(R^*,t)=J_D(R^*,t)$. To simplify, we define the scaled variables $\hat{r}=r/\lambda_e$ and $\hat{t}=t/\tau_\lambda$, we drop numerical factors of order unity and neglect the $r_p$ in $\delta$ since $R^*\gg r_p$. The flux equality then reduces to:
\begin{equation}
\label{pecletflux}
    \lambda_e \left[1+\frac{\hat{t} }{\hat{r}^3}\right]^{2/3}=r_p \left[ \erfc \left( {\frac{\hat{r}}{\sqrt{\hat{t}}}} \right)+ \frac{\hat{r}}{\sqrt{\hat{t}}}  ~ e^{ -\hat{r}^2/\hat{t}}\right].
\end{equation}
In terms of these rescaled variables, the solution derived in the Introduction is simply equivalent to a capture radius $\hat{r}=1$ with a cleanup time $\hat{t}=1$. For long times, the solution of this equation is $\hat{r}^*=\hat{t}^{\nicefrac{1}{3}} (\lambda_e/r_p)^{\nicefrac{1}{2}} $. Returning to our previous variables, this becomes
\begin{equation}
 \label{pecletfluxr*}
{R^*(t)} \approx  \lambda_e \times \left[t/\tau_\lambda\right]^{1/3} \times  (\lambda_e/r_p)^{1/2}.
\end{equation}
The additional $(\lambda_e/r_p)^{\nicefrac{1}{2}}$ term increases the value found in eq.~\ref{Driftdistance} by a factor $\approx 16$ for our numerical example. In 2D, a similar calculation gives
\begin{equation}
\label{2dfluxR*}
   R^*_{2d}\approx \lambda_e \times \left[t/\tau_\lambda\right]^{1/6};
\end{equation} 
the capture radius is thus predicted to increase more slowly in 2D and to reach $\lambda_e$ at $t=\tau_\lambda$.

\subsection{Reinterpreting the P\'eclet approach}

The approach described in the Introduction (eqs.~\ref{tauDscaling}-\ref{tauE}) used simple scaling arguments and compared two capture times for a single particle. In contrast, eq.~\ref{pecletfluxr*} gives the corresponding result when we consider particle fluxes, while eq.~\ref{Driftdistance} considers single particle dynamics but takes into account the small size of the target. 

However, there is yet another way to use a P\'eclet approach. Given a hemispherical region of radius $R^*$, the time needed for the electric forces to empty it is given by eq.~\ref{tauEro2r}: $\tau_E(R^*,0) \! = \! R^{*3}/3D \lambda_e$. To refill it by diffusion requires a time $\sim \! R^{*2}/D$ (this is similar to the fluorescence recovery after photobleaching (FRAP) of a spherical patch\cite{saxtonAnomalousSubdiffusionFluorescence2001}). These two times are equal at distance $R^*\approx \lambda_e$ and time $\approx \tau_\lambda$. This approach does not include the size of the target, but it may be misleading since the concentration outside the capture zone decreases with time, making refilling less efficient (this is why $R^*$ increases with time in eq.~\ref{pecletfluxr*}).

\section{Lattice Monte Carlo: Methods}
\label{section:LMCm}
We now use two-dimensional (2D) \textbf{L}attice \textbf{M}onte \textbf{C}arlo (LMC) simulations to study the capture of particles in the system shown in Fig.~\ref{Fig_system} and to investigate different ways to defining a capture radius or length scale (we chose 2D instead of 3D in order to be able to simulate larger systems and longer times; given the symmetry around the axis normal to the plane, we can easily infer 3D dynamics from our results). Since only the capture process is of interest here, the \textit{trans} side of the system is not part of the simulation.
\subsection{The simulation algorithm}
\label{section:LMC}
At each time step, the particle can move over a distance $\pm a$ (corresponding to the lattice step) in either the $x$ or the $z$ direction on the square lattice, or can remain at its initial position. The lattice step (and thus the "size" of the point-like particle) is $a\!=\!|z_{i \pm 1}\!-\!z_i|\!=\!|x_{j \pm 1}\!-\!x_j|\! =\!r_p/5$.

The trajectory of a Brownian particle can be discretized as a series of jumps on a lattice. For the current problem, building a LMC algorithm requires special care since the electric field is along a Cartesian axis and varies considerably between lattice sites. Moreover, as shown in ref.\cite{gauthierBuildingReliableLattice2004}, LMC algorithms often fail to give the right diffusion coefficient if the system is highly biased. Our approach will be to combine two 1D LMC algorithms (one for each Cartesian direction) into one, and then select the time step so that the resulting 2D LMC is reliable. In 1D, a particle on site $i$ (local potential energy $\Psi_i$) "jumps" to site $i\pm1$ at a rate
\begin{equation}
    \label{eq:prob}
    R_{\pm}={p_{\pm}}/{\tau_{\pm}},
\end{equation}
where $p_{\pm}$ is the probability for the particle to reach site $i\pm1$ before site $i\mp1$, and $\tau_{\pm}$ is the mean duration of this event. These probabilities and time durations are functions of the energy differences $\delta \Psi_{\pm}=\Psi_{i\pm1}-\Psi_{i}$ and are given by eqs. 3.20 - 3.22 in ref. \cite{hubertTheoreticalStudyPolymers2004a} (note the typo: $e^{u_-}$ must be replaced by $e^{u_\mp}$ in eq. 3.22) with $u_\pm=\delta \Psi_\pm/2$ and  $\tau_B=a^2/2D$. The probability of jumping from site $i$ to site $i \pm 1$ during a LMC time step $\delta t$ is thus $P^i_{\pm} = R^i_{\pm} \times \delta t$. The time step must satisfy the condition $\delta t \le \delta t_c \equiv min[1/(R^i_++R^i_-)]_i$ to ensure that $P^i_{ +}+P^i_{-} \le 1$ $~\forall i$; this implies a finite probability $P^i_o$ of not jumping during a time step on most sites $i$.

Using eq.~\ref{eq:sol}, we can write the particle's dimensionless potential energy at an oblate distance $\xi$ from the pore as
\begin{equation}
\label{eq: psimu}
\Psi(\xi) = \frac{QV(\xi)}{k_BT}=\frac{\lambda_e}{r_p}~\arctan \left[\sinh(\xi) \right].
\end{equation} 

In 2D, we use the same equations for each Cartesian direction, but the constrain on the time step now reads
\begin{equation}
\label{eq:criticalT}
    \delta t_c= min\left[ \frac{1}{\sum_{4_i} R^i}\right]_i,
\end{equation}
where $4_i$ means that the sum is over the 4 possible jump directions from site $i$.

This LMC algorithm can be used in two different ways:

I) To study the motion of a given particle: Using the probabilities described above, random numbers are generated at each time step to choose whether the particle will jump over a length $\pm a$ (the lattice step) in either the $x$ or $z$ direction, or stay put. This is the method used for Estimates 1 and 2 in the next section.

II) To follow a population of particles on the lattice: The time evolution of the particle concentration can be studied by iterating the master equation. The concentration $m_{x,z}^{j+1}$ at position ${x,z}$ and time step $j\!+\!1$ reads:
\begin{eqnarray}
{m_{x,z}^{j+1}}&=&P_{[x,z-1\rightarrow z]}m_{x,z-1}^{j}+P_{[x,z+1\rightarrow z]}m_{x,z+1}^{j}+\nonumber\\
&& P_{[x-1\rightarrow x,z]}m_{x-1,z}^{j}+P_{[x+1\rightarrow x,z]}m_{x+1,z}^{j}+\nonumber\\&&P_{[o]}  m_{x,z}^{j}~~,
\end{eqnarray}
where $x,z,j$ are integers and $P_{[o]}$ is the probability of not jumping. This was used in Estimates 3 and 4 below.

\section{Simulation Results and Analysis}
\label{Section:Results}

In this section, we present four different ways to analyze simulation data and define a length scale describing the capture process. The first two use single particle dynamics (allowing us to reach longer times), the third one examines the time-dependence of the capture rate, while the last one uses the concentration profile near the pore. 

\subsection{Estimate 1: 50$\%$ probability of capture}
\label{section:estimate_1}

Let $P(r,t)$ be the probability for a particle initially at radial position $r$ to reach the pore in a time $\le \! t$. Obviously, $P(r,t)$ increases with $t$ and decreases if the initial position $r$ is further from the pore. A straightforward length scale that can characterize the capture process is the distance $r \! = \! R^*_{\nicefrac{1}{2}}$ at which  $P \! = \! 1/2$. Since $R^*_{\nicefrac{1}{2}}$ is a function of time, we can use the solution of $P(R^*_{\nicefrac{1}{2}},t_{exp})=\nicefrac{1}{2}$ for a given experiment of duration $t_{exp}$.

\begin{figure}
\begin{center}
\includegraphics{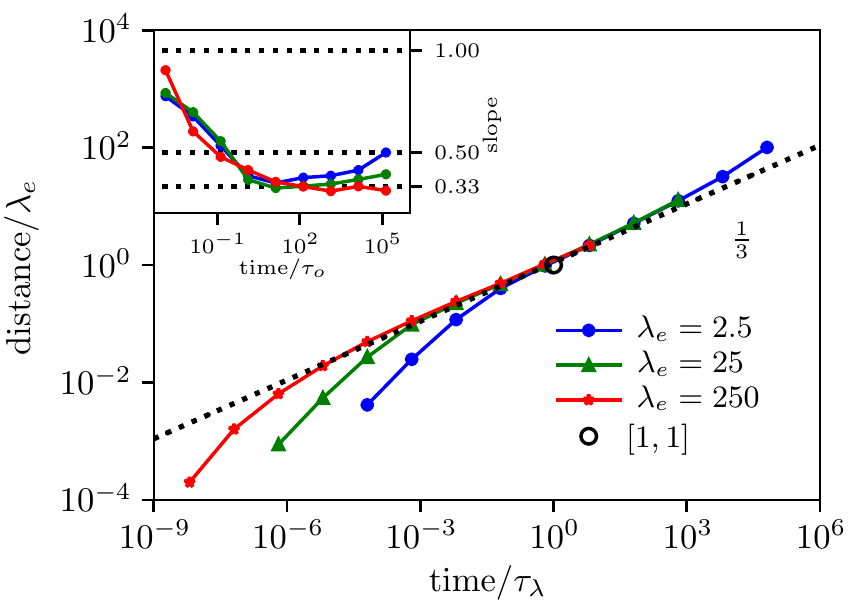}
\end{center}
\caption{Log-log plot of the capture length scale $R^*_{\nicefrac{1}{2}}(t)$ (in units of $\lambda_e$) \textit{vs.} time $t$ (in units of $\tau_\lambda$) for three electrostatic lengths $\lambda_e$. The empty circle marks the capture radius and time defined in eqs.~\ref{eq:R*2} and \ref{tauEL20}. An ensemble of 5000 particles were placed at various distances from the pore centre and the simulation was carried out for different time durations; the $y$-axis shows the distance at which $50\%$ of the particles were captured in the given time. The dash line has a slope of $\nicefrac{1}{3}$. Inset: Local slope \textit{vs.} time $t$ (in units of $\tau_o$). The key values of 1, $\nicefrac{1}{2}$ and $\nicefrac{1}{3}$ are marked by horizontal lines. }
\label{Fig:Ads}
\end{figure}

Figure \ref{Fig:Ads} shows the time evolution of $R^*_{\nicefrac{1}{2}}(t)$ in our 2D simulations for different field intensities (i.e., different electrostatic lengths $\lambda_e$) over a wide range of time scales; since there is a small angular dependence when eq.~\ref{eq: psimu} is used, all initial positions are chosen (for simplicity) to be along the vertical axis going through the pore centre (this has a small effect at short times). The simulation box is of size $R_b \! = \! 10^3~\lambda_e$ and its walls are reflecting: we thus have a finite loading box with no addition of analyte during the run. The $y$-axis is rescaled by the nominal capture radius $\lambda_e$ and the $x$-axis by the time $\tau_\lambda$. We first note that $R^*_{\nicefrac{1}{2}}(t \! = \! \tau_\lambda) \! \approx \! \lambda_e$ (the empty circle), in agreement with both eqs.~\ref{Driftdistance} and \ref{2dfluxR*}. Since $R^*_{\nicefrac{1}{2}}$ keeps increasing beyond that point, $\lambda_e$ is not sufficient to describe the nanopore capture properties. The data collapse for $t>\tau_\lambda$ is expected from our theoretical work.

In order to investigate the various regimes, the slope of the data is given in the Inset; note that the slope gives the local time exponent $\alpha$ if we assume a power law scaling $R^*_{\nicefrac{1}{2}}(t) \sim t^\alpha$. The pore region clean up time is given by eq.~\ref{eq:taurpovertaulambda}; for times shorter than this, the distances are comparable to the pore size $r_p$ itself and since the field is fairly flat near the pore, the slope $\alpha$ approaches unity. We then have a regime where $R^*_{\nicefrac{1}{2}}(t) \sim t^{\nicefrac{1}{3}}$, as predicted by eq.~\ref{Driftdistance}; the width of this regime increases for higher fields and basically vanishes at low field. There is no $t^{\nicefrac{1}{6}}$ regime, which rules out eq.~\ref{2dfluxR*}. The exponent slowly increases towards $\alpha \! = \! {\nicefrac{1}{2}}$ at longer times, a value consistent with a diffusion-limited process. Equation~\ref{eq:2DR*} predicted in that this latter regime would exist for distances larger than $R^*_{\nicefrac{1}{2}}/\lambda_e \sim (R_b/\lambda_e)^{2/3} \approx 100$, 
consistent with our data. It is important to note that this transition is \textit{not} expected in 3D (for these boundary conditions) because theory then predicts that $R^* > R_b$ -- see eq.~\ref{Pe3d}.

We conclude that $R^*_{\nicefrac{1}{2}}$ provides a good estimate of the revised P\'eclet capture radius, and that the latter increases well beyond the nominal value $\lambda_e$ during a run.

\subsection{Estimate 2: Time reversal and particle escape}
\label{section:estimate_2}
\begin{figure}
\begin{center}
\includegraphics{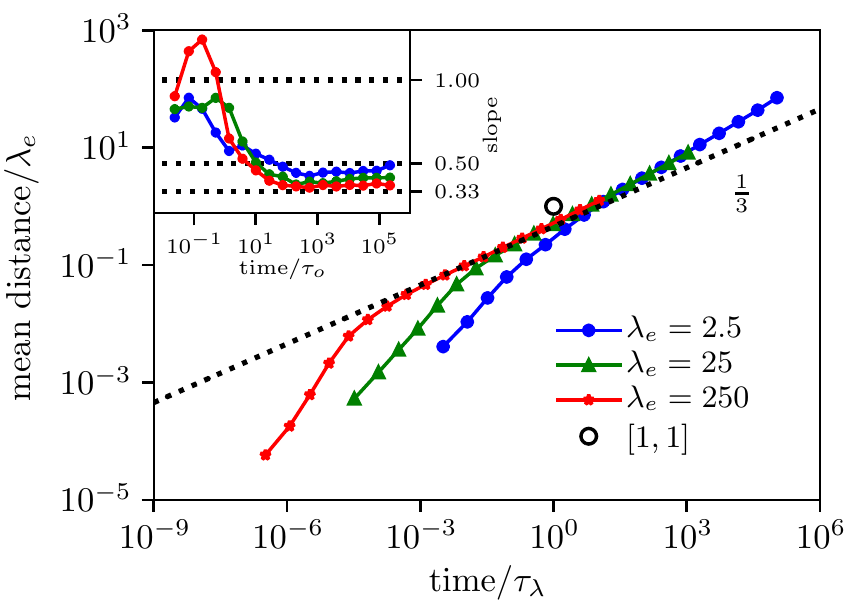}
\end{center}
\caption{Log-log plot of the mean distance $R^*_i$ (in units of $\lambda_e$)  reached by the particles at time $t$ (in units of $\tau_\lambda$) when the field polarity is reversed. An ensemble of 5000 particles were initially placed at the pore centre and mean displacements were computed at different times. The dash line has a slope of $\nicefrac{1}{3}$. Inset: The local slope \textit{vs.} time $t$ (in units of $\tau_o$). The key values of 1, $\nicefrac{1}{2}$ and $\nicefrac{1}{3}$ are marked by horizontal lines.}
\label{Dvstime}
\end{figure}

 One way to find a change in dynamics and estimate the corresponding length scale is to study the inverse process. Here we reverse the polarity of the electric field and let particles move away from the center of the pore (we now use an infinite box size). It is more efficient to track particles escape trajectories since we do not have to wait for diffusion to bring them into the capture zone. The mean displacement at time $t$ can be seen as a characteristic length scale for capture. The results shown in Fig.~\ref{Dvstime} agree with the trends seen in Fig.~\ref{Fig:Ads}, and the data sets again collapse passed the transient regime. The 'inverse capture radius' $R^*_i$, as defined here, increases during an experiment and $R^*_{i} \approx \lambda_e$ at $t=\tau_\lambda$ here as well. These results thus imply that $R^*_{\nicefrac{1}{2}} \propto R^*_i$ and confirm the existence of the three temporal regimes observed with Estimate 1.

\subsection{Estimate 3: The capture rate}
\label{section:estimate_3}

As discussed previously, the initial mean capture rate $\overline{\rho}_p$ is predicted to change with time because of the nature of the field very close to the pore (i.e., within a distance $\sim r_p$). More precisely, it is expected to increase with $t$ in 3D but to decrease with $t$ in 2D. We now use a reservoir-like boundary with a radius $R_r=10^4~r_p \gg \lambda_e$ in order to keep the concentration constant at $C(R_r,t) \! = \! C_o$ and study how the resulting capture rate evolves with time (we chose $R_r$ so the region close to $r \! = \! R_r$ remains essentially undisturbed during the simulation).

\begin{figure}
\begin{center}
\includegraphics{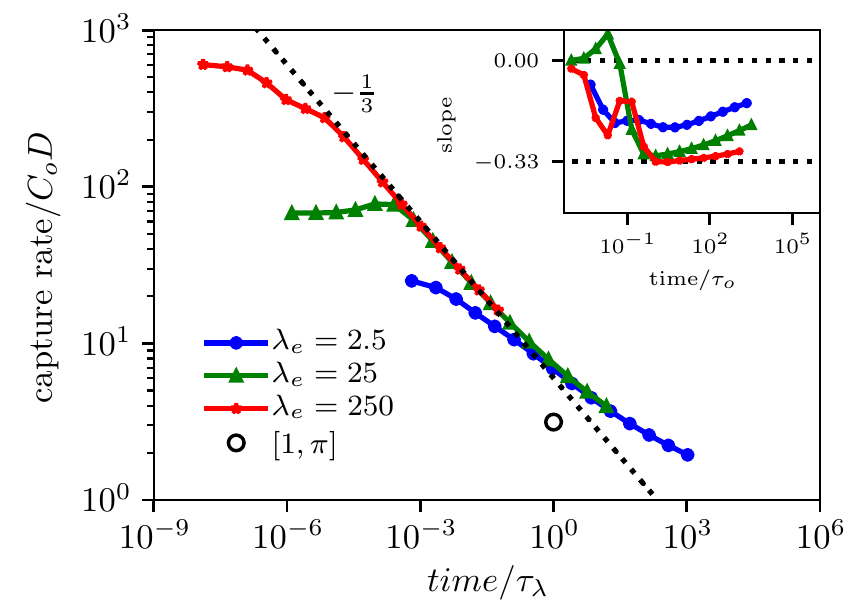}
\end{center}
\caption{Log-log plot of the capture rate ${\rho}(t)$ (in units of $C_oD$) \textit{vs.} time $t$ (in units of $\tau_\lambda$). The dash line has a slope of $-\nicefrac{1}{3}$. Inset: The local slope \textit{vs.} time $t$ (in units of $\tau_o$).}
\label{Flux_2D}
\end{figure}

Figure~\ref{Flux_2D} shows the time dependence of the pore capture rate ${\rho}(t)$. As predicted for 2D, $\rho(t)$ decays as $t^{-1/3}$ (see eq.~\ref{InstantRate2D}) after the short time plateau. The latter is predicted to be at $\overline{\rho}_p/DC_o \approx \frac{\pi}{2}\lambda_e/r_p$; our data is in decent agreement with this prediction (except at low field) even though we did not consider angular effects. We also predicted that $\rho(\tau_\lambda)/DC_o \approx \pi$ in 2D, independent of the field intensity; this explains the collapse of the data around $t=\tau_\lambda$, and the predicted value is in decent agreement with the data (empty circle).

We showed in the previous two sections that the $r \sim t^{\nicefrac{1}{3}}$ regime evolves into a $t^{\nicefrac{1}{2}}$ regime at longer time in 2D. If we use $r \sim t^{\nicefrac{1}{2}}$ in eq.~\ref{N2D}, we obtain $N(t) \sim t$, and hence $\rho = {\partial N(t)}/{\partial t} = cst$; the Inset of Fig.~\ref{Flux_2D} shows that the data are indeed (slowly) converging towards a plateau regime at long times $t \gg \tau_\lambda$ (and distances $\gg \lambda_e$). 

We also predicted a 3D constant capture rate $\rho \approx \overline{\rho}_\lambda(t)$ at long times. The only difference of substance between 2D and 3D is the extra $r(t)$ term in eq.~\ref{Eq:sumcapture} when compared to eq.~\ref{N2D}. We can thus infer 3D from the product $r(t) \times \rho(t)$ using 2D data. Since we obtained $r(t) \! \sim \! t^{1/3}$ and $\rho(t) \! \sim \! t^{-1/3}$ in 2D, a 3D simulation would indeed give a constant capture rate for $t > \tau_p$, as observed by Ref\cite{charronPreciseDNAConcentration2019}.  

A characteristic length scale can be estimated from capture rates: since $N(t) \! = \! \int_{0}^{t} \rho(t')dt'$ molecules have been captured by time $t$, we can use eq.~\ref{N2D} to obtain the related length scale $R^*_{cap}(t)=\left[\frac{2}{\pi}N(t)/C_o\right]^{1/2}$. In a $\rho(t) \sim t^{-\beta}$ regime, this would give $N(t) \sim t^{1-\beta}$ and hence $R^*_{cap} \sim t^{(1-\beta)/2}$. Since $\beta$ decreases from $\nicefrac{1}{3}$ to $\approx 0$ past the transients in Fig.~\ref{Flux_2D}, $R^*_{cap}$ will scale just like $R^*_{\nicefrac{1}{2}}$ and $R^*_i$ for longer time (shown in Fig.~\ref{Fig:Ads} and Fig.~\ref{Dvstime}).

\subsection{Estimate 4: Quasi-static concentration profiles}
\label{section:estimate_4}

\begin{figure}
\begin{center}
\includegraphics{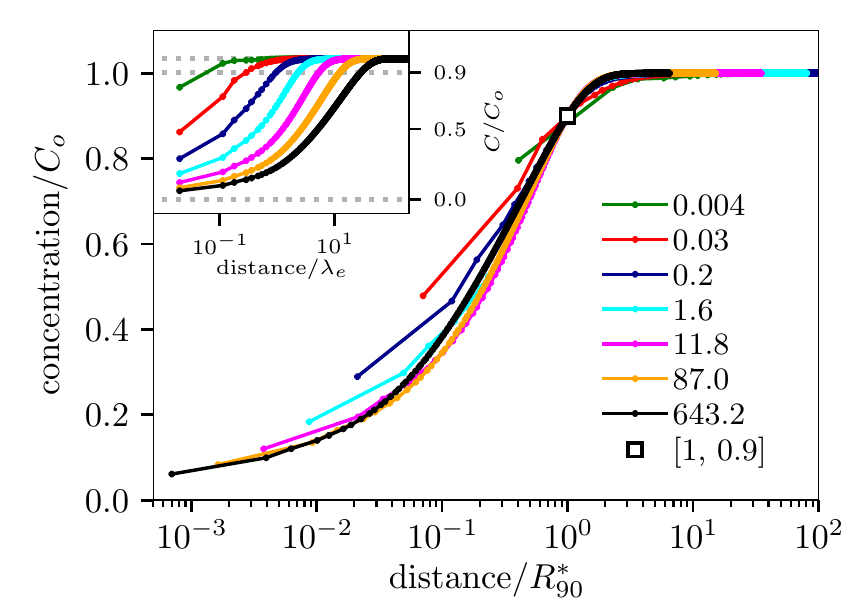}
\end{center}
\caption{Concentration \textit{vs.} distance from the pore (scaled by $R^*_{90}$, the distance at which $C= 0.9 ~C_o$) at different times (in units of $\tau_{\lambda}$) for $\lambda_e=2.5~r_p$. All the curves thus cross at the location of the empty square. The inset plot shows the concentration at different times \textit{vs.} distance in units of $\lambda_e$. } 
\label{Fig_CON}
\end{figure}

Keeping the boundary conditions used for Estimate 3, we now study the evolution of 2D concentration profiles -- Fig. \ref{Fig_CON}. Not surprisingly, a depletion region forms near the pore (Inset) and propagates outward: its width extends well beyond $\lambda_e$ and continues to grow for times $t \gg \tau_\lambda$. We define the width of the depletion region as the distance between the pore and the position $r=R^*_{90}$ where $C(R^*_{90})=0.90 ~C_o$. When the concentration profiles are plotted as a function of the rescaled distance $r/R^*_{90}$ (main part of the figure), the curves clearly evolve towards a steady-state distribution at long times.  

\begin{figure}
\begin{center}
\includegraphics{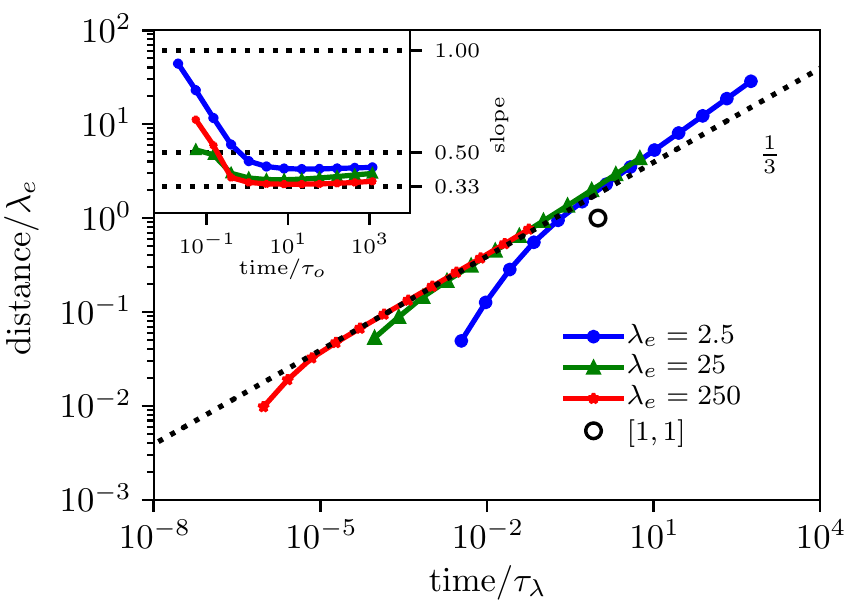}
\end{center}
\caption{Log-log plot of the depletion length scale $R^*_{90}$ (in units of $\lambda_e$) \textit{vs.} time (in units of $\tau_\lambda$). The dash line has a slope of $\nicefrac{1}{3}$. Inset: The local slope \textit{vs.} time $t$ (in units of $\tau_o$).}
\label{Fig_halfdrop}
\end{figure}

Figure~\ref{Fig_halfdrop} shows how the width $R^*_{90}$ of the depletion zone, which is also a length scale describing the capture process, increases with time. We again observe a $t^{\nicefrac{1}{3}}$ scaling after the initial transients; for longer times, our data is consistent with a diffusion-limited $t^{\nicefrac{1}{2}}$ regime.

\section{Discussion and Conclusion}
\label{section:conclu}

The concept of a capture radius $R^*$ that can characterize the capture process has been used in different ways in the literature. In this paper, we have revisited theoretical approaches often employed to define $R^*$. In particular, we did not work within the framework of a steady-state in order to see how relevant the length scale $R^*$ might be during an experiment. For readability reasons, Table ~\ref{tab:len&time} summarizes the definition of the important length and time scales used in this paper.

\begin{table}[h]
    \centering
\begin{tabular}{| M{1.3cm}||L{6.8cm}| }
 \hline 
 \bf{Symbol} & \bf{Definition} \\
 \hhline{|=||=|}
 $r_p$   & pore radius  \\
 \hline
 $r_e$ & characteristic length of the electrostatic potential outside the pore \\
 \hline
$\lambda_e$& traditional definition of the capture radius $R^*$ \\
\hline
$R_b$& distance between the outer reflecting boundary and the pore center \\
\hline
$R_r$& distance between the outer reservoir-like boundary and the pore  center\\
\hline
$R^*_{\nicefrac{1}{2}}$    & Estimate \#1 of the capture radius, defined as the location where the probability of capture is 1/2 \\
\hline
 $R^*_i$& Estimate \#2 of the capture radius, defined as the mean escape distance when the polarity is inverted\\
 \hline
  $R^*_{cap}$& Estimate \#3 of the capture radius, defined using the time dependent capture rate\\
 \hline
  $R^*_{90}$ & Estimate \#4 of the capture radius, defined as the width of the depletion zone where the concentration is 90\% of bulk \\
  \hline
   $\tau_{o}$& Basic unit of time :~ $\tau_o=r_p^2/D$ \\
   \hline
   $\tau_{exp}$ & Duration of an actual experiment \\
  \hline
 $\tau_{p}$& Time needed to empty a zone of radius $r_p$\\
 \hline
  $\tau_{\lambda}$& Time needed to empty a zone of radius $\lambda_e$ \\
 \hline
\end{tabular}
    \caption{Definition of the different length and time scales used in this paper.}
    \label{tab:len&time}
\end{table}

Our theoretical approach has been to use the P\'eclet number as a way to examine the nature of the capture process, thus treating nanopore capture as a separation process. In particular, we examined how the size of the target (i.e., the radius of the nanopore, $r_p$) can be an integral part of the calculation.

We first calculated the 'pure' diffusive and field-driven capture times for a particle in a finite size box with reflective boundary conditions. Our measure of the field intensity is the length scale $\lambda_e$, which is the capture radius predicted by previous theories (note that $\lambda_e \ne \lambda_e(r_p)$).  The resulting P\'eclet number predicts that $R^* \gg \lambda_e$ and even exceeds the box size. However, for realistic experimental times, $R^*$ simply grows as $t^{1/3}$ during the entire experiment (i.e., the capture remains field-driven throughout). A similar calculation in 2D indicates that it might be possible to transition into a diffusion-limited capture regime at long times. We also investigated a flux-based P\'eclet number using similar arguments. This approach predicted even larger capture radii but similar time-dependence.

In short, our P\'eclet calculations suggest that the conditions required to reach a diffusion-dominated capture regime might not be achievable in realistic situations: consequently, $R^*$ is simply a moving horizon ($\sim t^{1/3}$) that never reaches a steady state. 

To test the various concepts of a capture radius, we introduced a new Lattice Monte Carlo scheme to simulate (point-like) particle dynamics over a wide range of time scales and in the presence of a space-dependent field. The mesh size can be changed to explore finer short-time details or long time asymptotic limits. For instance, we were able to cover up to 7 decades in time with a fine mesh in 2D. Although this methodology can easily be extended to 3D, it would restrict us to rather small systems. This LMC algorithm can be used to look at both populations of particles and single particles with various boundary conditions.

\begin{figure}
\begin{center}
\includegraphics{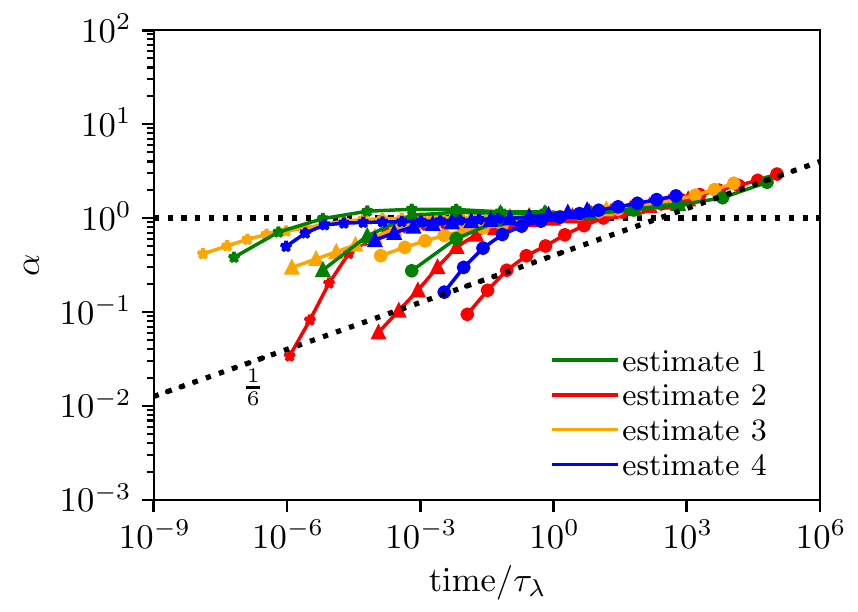}
\end{center}
\caption{The four different length scales (or capture radii $R^*$) introduced in the Results section \textit{vs.} time (in units of $\tau_\lambda$). The capture radii are rescaled by $R^*(t=\tau_\lambda,\lambda_e=25$) for each separate data set. Since the time evolution of the intermediate regime scales like $t^{1/3}$, the y-axis was further divided by $(t/\tau_\lambda)^{1/3}$ (hence the plateau for intermediate times). The resulting y-axis is thus $\alpha=({R^*(t)}/{R^*(\tau_{25}))/ (t/\tau_\lambda)^{1/3}}$. The dashed line with a slope $\nicefrac{1}{6}=\nicefrac{1}{2}-\nicefrac{1}{3}$ corresponds to the $t^{1/2}$ diffusion-limited regime expected at long times. Three different field intensities are shown for each of the four definitions of $R^*$.}
\label{Fig_R*rescaled}
\end{figure}

We introduced four different ways (which were based on: probability of capture, escape trajectories, capture rate and concentration profiles) to define a length scale related to the capture performance of a nanopore from the 2D simulation data, and we compared them to the theoretical approaches. Figure~\ref{Fig_R*rescaled} shows a compact view of our LMC data for the various time-dependent length scales $R^*(t)$. The short time behaviour of $R^*$ depends on the type of measurement made and the details of simulation protocol; nevertheless, it is clear that the fact that the field strongly deviates from the point-like approximation for distances $r<r_p$ leads to strong (but experimentally insignificant) short-time effects. For intermediate times, all estimates of $R^*$ increase like $t^{\nicefrac{1}{3}}$ (since the time axis has been rescaled by $t^{\nicefrac{1}{3}}$, this shows as a plateau on the figure). For much longer times, we observe a transition to a $t^{\nicefrac{1}{2}}$ regime, an indication that we have reached the diffusion-limited regime. This latter regime is occurring for distances much larger than the nominal capture radius $\lambda_e$ and for times much larger than the cleanup time $\tau_\lambda$. This is all consistent with our theoretical derivations. In particular, nothing special is observed at $\tau_\lambda$ or when the depletion regions reaches a distance $\lambda_e$. 
Except for the capture rate $\rho$, our 2D simulation results apply directly to a 3D system. The capture rate is predicted and observed to decrease very rapidly in 2D to reach a constant value only when (and if) the diffusion-limited capture regime is reached. In contrast, $\rho$ is going to quickly increase (because of the nature of the field near the pore) and reach a time-independent value in 3D. 
Therefore, we must conclude that $\lambda_e$ is simply a useful and natural way to measure the field intensity (for instance, the steady-state capture rate is linearly proportional to $\lambda_e$ -- see eq.~\ref{MeanRate}). Our revisited P\'eclet calculation indicates that one should observe a $R^*(t) \sim t^{\nicefrac{1}{3}}$ scaling law well beyond $\lambda_e$. However, this is actually a fairly slow growth process: in other words, the corresponding depletion region should remain in the $\sim 10~ \mu m$ range during a typical experiment, consistent with the results of Chen \textit{et al}. \cite{chenProbingSingleDNA2004}.

A real-life system may differ from the ideal conditions used here in different ways. For instance, we studied point-like particles in absence of EOF or hydrodynamic interactions with the walls. More importantly, such models assume that the system is closed, uniform and isolated (no flux through the system or thermal gradients): diffusion and field-driven motion are not necessarily the only two components to consider in a more general situation even in absence of EOF. However, a more complete theory that would include some of these effects for a particular system would nevertheless include the length and time scales derived here. Perhaps of more interest is the connection between the analyte transport properties (mobility and diffusivity), the local field intensity and the analyte conformation and/or orientation.  In the case of a flexible analyte such as DNA, for example, the molecule may deform in the gradient (moreover, its conformational entropy may act as a barrier). Similarly, if the analyte is not spherical, the field gradient will lead to orientation and thus to a position dependent diffusion coefficient $D(r)$; this latter case will be treated in a separate paper.



%
%

%

\begin{acknowledgements}
GWS acknowledges the support of both the University of Ottawa and the Natural Sciences and Engineering Research Council of Canada (NSERC), funding reference number RGPIN/046434-2013. LQ is supported by the Chinese Scholarship Council and the University of Ottawa. The authors wish to thank Vincent Tabard-Cossa, David Sean and Kyle Briggs for fruitful discussions. 
\end{acknowledgements}

\appendix
\section{The Electrostatic Field}
\label{app:Electric_field}
The field outside the pore is often modeled using the point-like approximation~\cite{wanunuElectrostaticFocusingUnlabelled2010,grosbergDNACaptureNanopore2010a,nomidisDNACaptureClyA2018a}, which greatly simplifies analytical calculations; for a pore of radius $r_p$ and length $l$, the corresponding potential $V(r)$ is the second equation in the Introduction. 
The part of the total potential gradient $\Delta V$ that is relevant for capture depends on the ratio of the access (ac) and channel (ch) resistances. The potential drop from infinity to the pore entrance is\cite{hallAccessResistanceSmall1975,kowalczykModelingConductanceDNA2011}
\begin{equation}
\label{seq:pdeop}
\delta V = \Delta V \times \frac{R_{ac}}{R_{ch}+2R_{ac}}=\Delta V \times \frac{\pi r_p}{ 4l+2\pi r_p},
\end{equation} 
where the resistances are $R_{ac} \! = \! {1/4r_p\sigma}$ and $R_{ch} \! = \! l/\pi r_p^2\sigma$ ($\sigma$ is the conductivity of the solution). If we take the potential to be zero at infinity, we obtain\cite{wanunuElectrostaticFocusingUnlabelled2010,grosbergDNACaptureNanopore2010a,nomidisDNACaptureClyA2018a}
\begin{equation}
\label{seq:EqpiontV}
V(r) = \frac{1}{2\pi \sigma r}\frac{\delta V }{R_{ac}}=\frac{2 r_p  }{\pi r} \delta V = \Delta V \frac{r_e}{r}
\end{equation}
where $r_e=r_p/(\frac{2l}{r_p}+\pi)$ is the characteristic length of the electrostatic potential outside the pore. The spherically symmetric electric field is then given by  
\begin{equation}
\label{seq:EqpiontE}
\overrightarrow{E}(r)= -\Delta V\frac{r_e}{ r^2} ~\hat{r}.
\end{equation}

Alternatively, we can use the exact solution of Laplace's equation in oblate spherical coordinates\cite{kowalczykModelingConductanceDNA2011,farahpourChainDeformationTranslocation2013a}, 
\begin{equation}
 \label{seq:sol}
  V(\xi,\eta,\phi) = \Delta V \frac{r_e }{r_p}\arctan \left[ \sinh(\xi) \right],
\end{equation}
where $\xi \in (- \infty, + \infty)$, $\eta \in [ 0, \pi ]$ and $\phi \in [ 0, 2 \pi ]$. The potential only depends on $\xi$ (here in units of $r_p$). The electric field can then be written as:
\begin{equation}
 \label{seq:Electricfield}
  \overrightarrow{E}(\xi,\eta) = -\frac{ \Delta Vr_e}{r_p^2 {\cosh\xi\sqrt{\sinh^2\xi+\sin^2\eta}}}\hat{\xi}.
\end{equation}
Expanding eq.~\ref{seq:Electricfield} for large oblate distances ($\xi \gg 1$) gives
\begin{equation}
    \label{Infinity}
   \overrightarrow{E}^{\infty} \approx -\frac{\Delta V r_e }{r_p^2\cosh \xi \sinh \xi}~\hat{\xi} \approx - \frac{\Delta V r_e}{r^2} ~\hat{r}, 
\end{equation}
in agreement with  \ref{seq:EqpiontE}.

Close to the pore ($\xi \rightarrow 0$), eq.~\ref{seq:Electricfield} reduces to 
\begin{equation}
    \label{eq:Electric_zero_ap}
   \overrightarrow{E}^{0} \approx -\frac{\Delta V ~ r_e}{r_p^2 \ } \times |\csc(\eta)| \times \left[1- \frac{\xi^2}{2} \left(1+\csc ^2(\eta)\right) \right] \hat{\xi}.
\end{equation} 
where $\eta \in [ 0, \pi ]$. The field is thus flat when $\xi\!<\!\xi^*(\eta)\!=\!\sqrt{2/(1+\csc ^2(\eta))}$. The critical length is $\xi^*(\pi/2)=1$ (or $r=r_p$) in the vertical direction and $\xi^*(0)=0$ along the wall (the field diverges at the pore edges).

\section{The mean first passage time}
\label{sec:diff_time}

We consider a particle initially placed between a small absorbing spherical target (at $r \! = \! r_D$) and a large reflecting spherical wall (at $r \! = \! R_b$). Its mean diffusive absorption time $\tau_D$ if it starts at radial position $R_b>r>r_D$ can be obtained analytically\cite{szaboFirstPassageTime1980a} from the equation 
\begin{equation}
  \frac{D}{r^{d-1}}~~ \frac{\mathrm{d}}{\mathrm{d}r}\left[r^{d-1}~\frac{\mathrm{d}\tau_D(r,r_D)}{\mathrm{d}r}\right]=-1,
\end{equation}
with the boundary conditions 
\begin{equation}
 \left[\tau_D(r,r_D)\right]_{r=r_{D}}=0;~\left[\tfrac{\mathrm{d}}{\mathrm{d}r}\tau_D(r,r_D)\right]_{r=R_b}=0,
\end{equation}
where $d$ is the dimensionality. In 2D we obtain 
\begin{equation}
\label{taud2D}
  \tau_{D}(r,r_D)=\frac{R_b^2}{4D}\left[2 \ln\left({\frac{r}{r_D}}\right) + \left(\frac{r_D^2-r^2}{R_b^2} \right) \right],
\end{equation}
while in 3D we recover 
\begin{equation}
  \tau_D(r,r_D)=\frac{R_b^2}{6D}\left[\frac{2R_b}{r_D}\left(1 \! - \! \frac{r_D}{r}\right) + {\left(\frac{r_D^2-r^2}{R_b^2} \right)} \right].
\end{equation}
As expected, $\tau_D$ diverges in both cases if the target size $r_D \rightarrow0$, hence the need to use a finite value for $r_D$, such as $r_D=r_p$ .

\bibliography{Refs}

\end{document}